\documentstyle{article}[12pt]
\newcommand{\be}{\begin{equation}}
\newcommand{\ee}{\end{equation}}

\def\bea{\begin{eqnarray}}
\def\eea{\end{eqnarray}}
\def\beas{\begin{eqnarray*}}
\def\eeas{\end{eqnarray*}}
\def\sla{\raise.15ex\hbox{$/$}\kern-.57em}

\def\part{\partial}
\begin{document}

\begin{center}
{\LARGE The Emergence of Anticommuting Coordinates  \\ and the 
Dirac-Ramond-Kostant operators}

\vspace*{10mm}

{\large Lars Brink}

{Institute of Theoretical Physics, Chalmers University of Technology}

{S-412 96 G\"oteborg, Sweden}
\end{center}

\section{Introduction}
Gunnar Nordstr{\"o}m died just a few years before the advent of quantum
mechanics. It is is very interesting, but of course futile to speculate how he
would have contributed to the new subject. His early attempts to increase the
number of space dimensions \cite{Nordstrom:fi} could very well have been very important
for him when the spin degrees of freedom were understood. The birth of 
quantum mechanics in the 1920's with the papers of
Heisenberg \cite{Heisenberg} and Schr\"odinger \cite{Schrodinger} and others 
really made a quantum jump in our
understanding of the microcosm. The formulation was revolutionary but 
treated space and time somewhat conservatively, since it was
non-relativistic and the newly discovered spin degrees of freedom
were just added on as multiplicative wave functions not connected to 
the $3$-dimensional space. In the hectic year of 1926 Klein \cite{Klein:fj},
Gordon \cite{Gordon} 
and Schr\"odinger \cite{Schrodinger} discussed the relativistic version and 
the Klein-
Gordon equation was suggested.

\be
((-i \hbar \frac{\partial}{\partial x^\mu})^2 + m^2)\Psi =0.
\ee

Schr\"odinger originally discarded this equation since it does not
give the correct spectrum for the hydrogen atom. In order to remedy
this problem, Darwin aand Pauli \cite{Pauli} added spin terms and succeeded to give 
a good but not perfect match of the spectrum.

Only Dirac seemed to worry about this problem. He talked to Bohr
about it when he visited Copenhagen, but Bohr assured him that the
problem would be solved within the existing formalism. Dirac had
other problems with the Klein-Gordon equation. He found that there is
non-positive probability following from the equation. Consider the
conserved current

\bea J^0 = &-i (\Psi^* \frac{\part}{\part t}\Psi - \frac{\part}{\part
t}\Psi^*~~\Psi) \\
J^i = &-i (\Psi^* \frac{\part}{\part x^i}\Psi - \frac{\part}{\part
x^i}\Psi^*~~\Psi).
\eea

Dirac wanted to have a probability $|\Psi|^2$. Then one sacred
evening in front of the fireplace at St John's College he found the
answer to his problems. The equation must be linear in $\frac{\part}{\part
t}$ and he was led to the equation that would once be on his tombstone
in Westminster Abbey \cite{Dirac:1928hu}.

\be
i \hbar \frac{\part \Psi}{\part t}= (-i \hbar
\overrightarrow{\alpha}\cdot\overrightarrow{\partial} + \alpha_{4}m)\Psi 
\ee
or in more modern form
\be
(i\gamma_{\mu}\part^{\mu}-m)\Psi=0.
\ee

For this equation Dirac found a positive probability through the
conserved current

\bea &J^0 =  \bar\Psi \gamma^0\Psi=|\Psi|^2\\
&J^i = -\bar\Psi \gamma^i\Psi.
\eea

There was a phenomenal success for the equation. It really gave the
correct spectrum for the hydrogen atom taking spin into account.
However, as it happens for some of the most gifted scientists, the
result was correct but the starting point not fully correct. With the
understanding of the positron solution one understood that the Dirac
current should be multiplied by the charge and should be interpreted 
as a current for the charge leading to positive or negative charge
densities depending on the sign of the charges present. Similarly the
Klein-Gordon equation was understood to describe scalar particles and
again the current should be understood as a charge current. It is
fair to say that the Dirac equation was the starting point for the
Standard Model. The nonconservation of particles led to quantum field
theory and eventually it was realized that such theories need
non-abelian gauge fields for $d=4$.

There was one observation in Dirac's paper that never caught on at
the time. It is natural since the equation worked so well so one did 
not need to theorize over the origin of the equation. Dirac made a
point that the $\gamma$'s are dynamical and on an equal footing to
$x^\mu$ and $p^\mu$. That point was dormant for more than forty years
until Ramond utilized this fact when he discovered superstring theory 
\cite{Pierre}.

There were other issues that nobody really raised. The algebra
between $\gamma \cdot p$ and $p^2$ is a graded algebra and this was
in fact the first graded algebra in physics, but its mathematical
properties were not really exploited. Neither did one discuss the
obvious graded algebra between Dirac fermions and Klein-Gordon
bosons, which is the basis for supersymmetry. The use of Grassmann
variables was not understood until fermions were used in Lagrangian
formulations of quantum field theories. All these properties came up 
in the 1970's when we tried to find new models. In Dirac's time he
saw no need for new models. The equation worked so beautifully.

One can speculate what would have happened if Dirac had stayed in
Bristol and had been shielded from the events of 1925 and 1926.
Perhaps he would have discovered his formalism from 1949 first \cite{Dirac:cp}.
In that 
paper Dirac showed that any direction inside the light-cone including 
the limiting light-cone itself can be used as the evolution parameter 
for the relativistic problem. Specifically one can use 
$x^+=\frac{1}{\sqrt2}(x^0+x^{d-1})$. From the mass-shell relation

\be
p^2+m^2= -2p^+p^-+ {p^i}^2 +m^2=0,
\ee
one can solve for

\be
p^-= \frac{1}{2p^+}({p^i}^2 +m^2).
\ee

This equation is linear in the light-cone energy $p^-$ and the 
corresponding wave equation reads

\be
-i\hbar \part^- \Psi= (-{\hbar}^2 \frac{1}{2p^+} {\part^i}^2 + 
\frac{m^2}{2p^+}) \Psi.
\ee

The equation is identical for spin-$0$ and spin-$1/2$ particles. 
They differ in their Grassmann properties of the $\Psi$ only. However, 
on our planet the development went the other way.

\section{Dual Resonance Models}
String theory was born in 1968 with the celebrated paper by Veneziano
\cite{Veneziano:yb}. 
He constructed a four-point amplitude with magical properties. In 1969 
Koba and Nielsen \cite{Koba:1969rw} found the N-point generalization of this
amplitude 
which subsequently was factorized by Fubini and Veneziano \cite{Fubini:1969qb}
and by Nambu \cite{Nambu}. 
The important constructs are 

\be
Q^{\mu}(z)= x^{\mu} - i p^{\mu}lnz - i \sum_{n=1}^{\infty} 
\frac{1}{\sqrt{n}}({{a^\dagger}_{n}}^{\mu} z^n-{a_{n}}^{\mu}z^{-\mu}).
\ee

\be
P^{\mu}= iz \frac{dQ^{\mu}}{dz},
\ee
which generalize the usual phase space of a point-particle. 
Since these amplitudes describe open strings in modern terms, the world 
sheet variable $z=exp(i\tau)$. In Koba-Nielsen's work it was just an 
internal parameter. The amplitudes were very interesting but they 
lacked fermions.

On January 4, 1971 Pierre Ramond \cite{Pierre} submitted a paper which was the 
starting point for the Superstring Theory. In this paper he used the 
insight that Dirac had back in 1926 and put it into practise into 
dual resonance models, the name that the subject had at the time. He 
realized that the $x$ and the $p$ in the preceeding formulae are zero 
modes and he devised an averaging procedure which is like taking the 
zero fourier mode, where

\bea
x^{\mu}=&<Q^{\mu}(z)>,
p^{\mu}=&<P^{\mu}(z)>.
\eea
and a correspondence principle for how to go from the usual field 
theory equations (Klein-Gordon) to the dual resonance models equations

\bea
&p^2+m^2  =  <P^{\mu}><P_{\mu}> +m^2 \Rightarrow &\\
&(<P^{\mu}P_{\mu}> +m^2 )\;| phys>   =  0&,§
\eea
or 

\be
(L_{0}-m^2)| phys> =0.
\ee
        
Similarly the Virasoro operators \cite{Virasoro:1969zu} are given by

\be
L_{n}= <z^n\;P^2(z)>.
\ee

In this formalism it is natural to add in fermions. Remember that 
Dirac said that the $\gamma$'s should be regarded as dynamical 
variables. We can now use the correspondence principle and introduce 
a $\Gamma^{\mu}(z)$ such that

\be
\gamma^{\mu}= <\Gamma^{\mu}(z)>
\ee

with the natural anticommutation relation

\be
\{\Gamma^{\mu}(z),\Gamma^{\nu}(z')\}=2\eta^{\mu\nu}\delta (z-z').
\ee

A solution to this is 

\be
\Gamma^{\mu}(z) = \gamma^{\mu} + i \sqrt{2} \gamma^5 
{\sum_{n=1}}^\infty ({{b^\dagger}_{n}}^{\mu} z^n +{b_{n}}^{\mu}z^{-n}).
\ee

It is now rather natural to try to generalize the Dirac equation in 
the same way the Klein-Gordon equation led to the Virasoro constraints.

\bea
&\gamma\cdot p-m= <\Gamma_{\mu}(z)><P^{\mu}(z)> - m=0 
\Rightarrow& \\
&[<\Gamma_{\mu}(z) P^{\mu}(z)> - m]\;|phys>=0.&
\eea

Square the two expressions.

\bea
&<\Gamma \cdot P><\Gamma \cdot P> - m^2\;=0 \Rightarrow&\\
&[<P^2(z)>-i/4<\Gamma(z) z\frac{d}{dz} \Gamma(z)> -m^2]\;|phys>=0.&
\eea

This is the new mass shell condition $L_{0} - m^2 =0$ and the fourier 
modes are the new Virasoro operators

\be
L_{n}= <z^n P^2(z)>-i/4<z^{n+1}\Gamma(z) z\frac{d}{dz} \Gamma(z)>. 
\ee
Similarly the Dirac equation led to the constraint operators

\be
F_{n}= <z^n \Gamma_{\mu}(z) P^{\mu}(z)>,
\ee
which together with the new $L_{n}$'s give the superVirasoro algebra and 
hence opened the way to 
construct scattering amplitudes involving propagating fermions.

This was the first example of a 2-dimensional superconformal algebra 
extending a 2-dimensional supersymmetry and it led eventually to 
4-dimensional supersymmetry \cite{Wess:tw}. The rest is history.

\section{Anticommuting coordinates}

It is clear from the discussion above that the $\gamma$-matrices and 
their extensions should be interpreted as being dynamical coordinates 
carrying degrees 
of freedom. If one considers the classical action for a scalar 
particle and the one for a spinning one, the relation between the 
usual coordinates and the fermionic ones become clear.

The usual action for a spinless particle 

\be
S=m \int d\tau \sqrt{-{\dot x}^2}
\ee
can be rewritten as 

\be
S=\frac{1}{2}\int d\tau (e{\dot x}^2 - \frac{1}{e} m^2).
\ee

The function $e$ is an einbein. The latter form can be used for the massless case. This expression 
lends itself to a generalization to include Grassmann odd 
coordinates $\psi^{\mu}$ \cite{Brink:1976sz}.

\be
\label{S}
S=\frac{1}{2}\int d\tau (e{\dot x}^2 + \psi\cdot \dot\psi + \lambda 
\dot x \psi).
\ee

In here $\lambda$ is a superpartner to the einbein $e$. There is a 
(necessary) difference between the bosonic coordinate $x$ and the 
fermionic one $\psi$ in that the latter occurs with just one 
derivative in the action. This leads to a constrained phase space 
with no momentum conjugate to $\psi$. (We can also write it as $d/2$ 
coordinates and $d/2$ momenta.)

In the quantum case the usual representation is 

\be
\psi^{\mu} \rightarrow \; \gamma^{\mu}.
\ee

We can also introduce a more coordinate-like  representation by 
introducing (in the $4$-dimensional case as an example)
two Grassmann odd coordinates $\theta^1$ and $\theta^2$ 
\cite{Brink:1975qk} and write

\bea
\psi^0 + \psi^3 &=& \sqrt{2}\;\theta^1,\\
\psi^1 + i \psi^2&=&Ê\sqrt{2}\;\theta^2,\\
\psi^0 - \psi^3 &=& \sqrt{2}\;\frac{d}{d\theta^1},\\
\psi^1 - i \psi^2&=&\sqrt{2}\;\frac{d}{d\theta^2}.
\eea

The corresponding field will then be a function of $x^{\mu},\; 
\theta^1, \;\theta^2$, ie have four components like a spinor.

Another type of fermionic coordinate is the spinorial one $\theta^a$, 
which was pioneered by Montonen \cite{Montonen} in string theory and then taken 
over 
to 4 dimensions by Salam and Strathdee \cite{Salam:1974jj}. We get that if we 
instead of 
the vectorial fermionic coordinate in (\ref{S}) use a spinorial one 
and change the action to the action for the superparticle \cite{Casalbuoni:1976tz}.

\section{What the Physicists did not do}
Consider again the Dirac operator $ {\not{\!\!P}}=\gamma_{\mu}P^{\mu}$. We can 
regard the $P^{\mu}$'s as generators for the translation algebra. The 
Dirac operator satisfies the algebra

\be
\{{\not{\!\!P}},\;{\not{\!\!P}}\}= P^2.
\ee

If we let $P^2$ act on a state 

\be
P^2 \Psi =0,
\ee
the result is nontrivial since the translation algebra is non-compact. 
What happens if we generalize this procedure to a Lie algebra?

\be
[T^a, \;T^b] = i f^{abc} T^c, \;\;\; a= 1, 2, \ldots D.
\ee

Define the Kostant-Dirac operator \cite{KOS} as 

\be
{\not{\!\!K}}= \gamma^aT^a- \frac{i}{2\cdot 3!}\gamma^{abc} f^{abc}
\ee
with $\{\gamma^a, \gamma^b\}=2\delta^{ab}$ and $\gamma^{abc}$, the 
three times antisymmetrized product of three $\gamma$'s.
Then 
\be
\{{\not{\!\!K}}, \;{\not{\!\!K}}\}= 2 C_{2} +\frac{D}{12} 
{C_{2}}^{adjt},
\ee
where $C_{2}$ is the Casimir operator of the representation of $T^a$ 
and ${C_{2}}^{adjt}$ is the Casimir operator of the adjoint 
representation. If we consider the Kostant-Dirac equation

\be
{\not{\!\!K}}\Psi= 0,
\ee
one finds that it has only trivial solutions for a compact algebra.

Consider instead Lie algebra cosets. Let
\vspace{.5cm}

$g$ be generated by $T^a$  $a=1\ldots D$, 

$h$ be generated by $T^i$  $i=1\ldots d$,

$g/h$ be generated by $T^m$  $m= 1\ldots D-d$.
\vspace{.5cm}

Consider so the Kostant operator for the coset $g/h$.

\be
{\not{\!\!K}}= \gamma^mT^m- \frac{i}{2\cdot 3!}\gamma^{mnp} f^{mnp}.
\ee

Now the square of the operator is 

\be
{\not{\!\!K}}^2= [C_{2}(g) +\frac{D}{24} {C_{2}}^{adjt}(g)] - [C_{2}(h) +
\frac{D}{24} {C_{2}}^{adjt}(h)],
\ee
where $h$ is generated by $L^i= T^i+S^i$ and $S^i= -\frac{i}{4} 
\gamma^{mn}f^{imn}$. This formula is remarkable in that the spin 
operators for the subalgebra use the structure constants of the full 
algebra.

The full generator $L^i$ of the subalgebra commutes with the Kostant 
operator ${\not{\!\!K}}$ and  hence the solutions to the the Kostant 
equation

\be
{\not{\!\!K}}\Psi= 0
\ee
have solutions assembled in full multiplets of $h$.

Take as an example $g$ to be the Poincar\'{e} algebra and $h$ to be the 
Lorentz one. The Lorentz algebra is then generated by 

\be
M^{\mu\nu}= T^{\mu\nu}+S^{\mu\nu},
\ee
where

\bea
S^{\mu\nu} &=& \frac{i}{4}\gamma^{\rho\sigma} f^{(\mu\nu)\rho\sigma}\\
&=& -\frac{i}{2}\gamma^{\mu\nu} ,
\eea
just as Dirac once told us.

If $g$ and $h$ have the same rank the solution to Kostant's equation 
can be written as \cite{Gross:1998bd}

\be
\Psi = V_{\lambda}\otimes S^+ -  V_{\lambda}\otimes S^-,
\ee
where $S^{±}$ are the two spinor representation of $SO(D-d)$ and 
$V_{\lambda}$ is the representation of $g$ of highest weight 
$\lambda$. The solution is an infinite sequence of $r$-tuplets with 
$r$ the Euler number if $g/h$. In some cases, the r-tuplets, or Euler multiplets,
have the same number of bosons and
fermions \cite{Pengpan:1998qn}, but no apparent supersymmetry, which still
make them very interesting from a quantum mechanical point of view.
Indeed, it has been shown that
\be
V_\lambda\otimes S^+ - V_\lambda\otimes S^-=\sum_{c}^r (-1)^w~U^{}_{c\bullet
\lambda}\ ,\ee
where the sum is over the $r$ elements of the Weyl group of $\bf g$ that
are not in $\bf h$'s, and $U_{c\bullet
\lambda}$ denote representations of $\bf h$, and the $\bullet$ denotes a 
specific construction which has been defined in \cite{Gross:1998bd}.

\subsection{Supersymmetric Euler Multiplets}
It is well-known that the degrees of freedom of supersymmetric
theories can be labelled in terms of the Wigner little group of the
associated Poincar\'e algebra. Since theories with gravity cannot sustain a
finite number of {\it massless} degrees of freedom with spin higher than two, the cosets
cannot exceed sixteen dimensions, which yield the two  spinor representations of $SO(16)$,
 with 256 degrees of freedom.

We have found \cite{Brink:1999te} only a few cosets, with $D-d=16,8,4$, for which the basic
Euler multiplets have the right quantum numbers to represent massless particles of relativistic theories:

\vskip .5cm
\noindent $\triangleright$ {\bf $16$-dimensional Cosets}
\vskip .3cm
\noindent The $256$ states of the associated Clifford  are generated by the
two spinor irreps of
$SO(16)$, yielding three possible theories:
\begin{itemize}
\item $SU(9)\supset SU(8)\times U(1)$
with lowest Euler multiplet
$$ {\bf 1}_2 \oplus {\bf 8}_{3/2}\oplus  {\bf 28}_1 \oplus {\bf 56
}_{1/2}\oplus {\bf 70}_0 \oplus
{\overline {\bf 56}}_{-1/2} \oplus {\overline {\bf 28}}_{-1}\oplus
{\overline {\bf 8}}_{-3/2}
\oplus {\bf 1}_{-2} $$ Interpreting the $U(1)$ as the helicity little group
in four dimensions,
these can be thought of as  the massless spectrum of either type IIB string
theory,  or  of
$N=8$ supergravity.
\item $SO(10)\supset SO(8)\times SO(2)$
with lowest Euler multiplet
$$ {\bf 1}_{2}\oplus  {\bf 8}_{3/2}\oplus  {\bf 28}_{1} \oplus {\bf
56}_{1/2} \oplus ({\bf 35}_{0}
\oplus {\bf 35}_{0}) \oplus {\bf 56}_{-1/2} \oplus {\bf 28}_{-1}\oplus
{\bf 8}_{-3/2}\oplus  {\bf
1}_{-2} $$

Viewing $SO(8)$ as the little group in ten dimensions, these may represent
the massless spectrum of
IIB superstring in ten dimensions, but only after using $SO(8)$ triality.
Alternatively, with
$U(1)$ as the helicity, it could just be the massless particle content of the
type IIB superstring
theory, or of $N=8$ supergravity in four dimensions.
\item $SU(6)\supset SO(6)\times SO(3)\times SO(2)$
with lowest Euler multiplet
$$({\bf 1},{\bf 1})_{2}\oplus ({\bf 4},{\bf 2})_{3/2}\oplus ({\bf
10},{\bf 1})_1
\oplus ({\bf 6},{\bf 3})_1\oplus ({\bf 4},{\bf 4})_{1/2}\oplus ({\bf
20},{\bf 2})_{1/2}$$
$$~~~~~~~~~~~\oplus ({\bf 20^\prime},{\bf 1})_0\oplus ({\bf
15},{\bf 3})_0
\oplus ({\bf 1},{\bf 5})_0$$
$$
\oplus ({\bf 4},{\bf 4})_{-1/2}\oplus ({\overline{\bf 20}},{\bf 2})_{-1/2}
\oplus ({\overline{\bf
10}},{\bf 1})_{-1}
\oplus ({\bf 6},{\bf 3})_{-1}\oplus ({\overline{\bf 4}},{\bf 2})_{-3/2}
\oplus({\bf 1},{\bf 1})_{-2}\ .$$
This is the massless spectrum of $N=8$ supergravity, or of type IIB superstring
in four dimensions, or of massless theories in five and eight dimensions.

\item $F_4\supset SO(9)$
The lowest Euler multiplet
$$ {\bf 44}\oplus  {\bf 84}\oplus  {\bf 128} $$
describes the massless spectrum of $N=1$ supergravity in $11$ dimensions,
the local limit of M-theory, by identifying $SO(9)$ as the massless little
group.

\end{itemize}
\vskip .3cm
\noindent $\triangleright$ {\bf $8$-dimensional Cosets}
\vskip .3cm
\noindent The Clifford algebra is realized on the $16$ states that span the
two spinor irreps of
 $SO(8)$, leading to the two massless interpretations
\begin{itemize}
\item $SU(5)\supset SU(4)\times U(1)$
 which yields the  lowest Euler multiplet
$$ {\bf 1}_1\oplus {\bf 4}_{1/2} \oplus  {\bf 6}_0 \oplus  {\overline {\bf
4}}_{-1/2} \oplus
{\bf 1}_{-1}$$
With $U(1)$ as the helicity little group, this particle content is the same
as the massless
spectrum of $N=4$ Yang-Mills in four dimensions.

It could also describe a massless theory in eight dimensions with one
conjugate spinor, and eight
scalars, since $SU(4)\sim SO(6)$.

\item $SO(6)\supset SO(4)\times SO(2)$
with lowest Euler multiplet
$$ ({\bf 1},{\bf 1})_1\oplus  ({\bf 2},{\bf 2})_{1/2} \oplus ({\bf 1},{\bf
3})_0\oplus  ({\bf 3},
{\bf 1})_0\oplus  ({\bf 2},{\bf 2})_{-1/2}\oplus  ({\bf 1},{\bf 1})_{-1}$$
This particle content is the same as the $N=4$ Yang-Mills in four dimensions.

\end{itemize}
\vskip .3cm

\vskip .3cm
\noindent $\triangleright$ {\bf $4$-dimensional Cosets}
\vskip .3cm
\noindent The Clifford algebra is realized on $4$ states of the two spinor
irreps of $SO(4)$.
They appear as the lowest Euler multiplet in the following decomposition:
\begin{itemize}
\item $SU(3)\supset SU(2)\times U(1)$.
The lowest Euler multiplet$$ {\bf 1}_{1/2}\oplus {\bf 2}_{0}\oplus {\bf
1}_{-1/2}$$
can be interpreted as a relativistic theory in $4$ dimensions, since the
$U(1)$ charges are half
integers and integers, and it can describe the massless $N=1$ Wess-Zumino
supermultiplet in four
dimensions, as well as a massless supermultiplet in five dimensions.
\end{itemize}

So far, we have limited our discussion to massless degrees of freedom, but
we should note for
completeness that possible relativistic theories with  massive degrees of
freedom appear in the
cosets $Sp(2P+2)\supset Sp(2P)\times Sp(2)$ with the same content as
massive $N=P$ supersymmetry in
four dimensions. The massive $N=4$ supermultiplet in four dimensions is
also described by the coset
$SO(8)\supset SO(4)\times SO(4)$, and the massive $N=2$ supermultiplet in
four dimensions can be
described in terms of $G_2/SU(2)\times SU(2)$.

In this section, we have shown that the degrees of freedom of some
supersymmetric theories are nothing but solutions of the Kostant-Dirac
equation associated with specific  cosets. 
Furthermore, not all supersymmetric theories appear in this
list, only the local limit of M-theory, of type
IIB superstring theories (not type IIA, type I,
nor heterotic superstrings), and certain local field theories, $N=4$
Yang-Mills, and
$N=1$ Wess-Zumino multiplets. All the massless cosets are both hermitian
and symmetric, except
for the Wess-Zumino multiplet and $N=1$ supergravity in eleven dimensions
which are only symmetric.

\section{An Extension of 11-dimensional Supergravity}
As we saw above, one of multiplets to come out of Kostant's equation is the
$11$-dimensional supergravity one. Since it comes in the form of 
the $SO(9)$ multiplet, it is natural to consider the light-cone 
formulation of this model. The dimensionally reduced $4$-dimensional 
$N=8$ supergravity was constructed in this formulation up to the 
three-point function some twenty years ago by Bengtsson, Bengtsson 
and Brink \cite{Bengtsson:1983pg}. Recently this action has been ``oxidized'' 
to 11 
dimensions \cite{Sud}. This is now a perfect starting point to attempt an 
extension to a model that contains all the Euler multiplets. The 
corresponding free such theory was constructed by Brink, Ramond and 
Xiong \cite{Brink:2002zq}.

The supergravity theory is described by a superfield in terms of an 
anticommuting coordinate $\theta^a$ which is an ${\bf 8}$ under $SO(7)$. 
Following the highest weight representations of the irreps in the 
multiplet we can write the superfield as 

\be
\Phi~=~\theta^1\theta^8\,h(y^-,\vec x)~+~\theta^1\theta^4\theta^8\,
\psi(y^-,\vec x)~+~\theta^1\theta^4\theta^5\theta^8\,A(y^-,\vec x)\ + 
lower \;\; weights ,
\ee
with
$y^-~=~x^--\frac{i\theta\overline\theta}{\sqrt{2}}\ .$
We now like to add internal bosonic degrees of freedom and let 
Kostant's operator work on the field. We then have to find a 
representation of the coset. We start by finding an oscillator 
representation of $F_{4}$ In this  case, it turns out that all representations 
of the exceptional group $F_4$ are generated by three (not four \cite{Fulton:1984gk}) 
sets of oscillators transforming as  ${\bf 26}$.

We label each copy of $26$ oscillators
as $A^{[\kappa]}_0,\; A^{[\kappa]}_i,\; i=1,\cdots,9,\; B^{[\kappa]}_a,\; 
a=1,\cdots,16$, 
and their hermitian conjugates, and where $\kappa=1,2,3 $. 
Under $SO(9)$, the $A^{[\kappa]}_i$ transform as ${\bf 9}$, $B^{[\kappa]}_a$ 
transform as ${\bf 16}$, 
and $A^{[\kappa]}_0$ is a scalar. They satisfy the commutation relations of 
ordinary   harmonic oscillators

\be
[\,A^{[\kappa]}_i\,,\,A^{[\kappa']\,\dagger}_j\, ]~=~\delta^{}_{ij}\,\delta_{}^
{[\kappa]\,[\kappa']}\ ,\qquad [\,A^{[\kappa]}_0\,,\,A^{[\kappa']\,\dagger}_0\, 
]~=~\delta_{}^{[\kappa\,\kappa']}\ . 
\ee
Note that the  $SO(9)$  spinor operators  satisfy Bose-like commutation
relations

\be[\,B^{[\kappa]}_a\,,\,B^{[\kappa']\,\dagger}_b\, ]~=~\delta^{}_{ab}\,
\delta_{}^{[\kappa]\,[\kappa']}\ .
\ee
The generators $T_{ij}$ and $T_a$  

\bea
T^{}_{ij}&=&-i\sum_{\kappa=1}^4\left\{\left(A^{[\kappa]\dag}_iA^{[\kappa]}_j-
A^{[\kappa]\dag]}_jA^{[\kappa]}_i\right)+\frac 12\,B^{[\kappa]\dag}\,
\gamma^{}_{ij} B^{[\kappa]}\label{t_ij}\right\}\nonumber\ ,\\
T_a&=&-\frac{i}{{2}}\sum_{\kappa=1}^4\left\{ (\gamma_i)^{ab}\left(A^{[\kappa]
\dag}_iB^{[\kappa]}_b-B^{[\kappa]\dag}_bA^{[\kappa]}_i\right)-\sqrt{3}\left(B^{
[\kappa]\dag}_aA^{[\kappa]}_0-A^{[\kappa]\dag}_0B^{[\kappa]}_a\right)\right\}
\nonumber\ ,\label{t_a}
\eea
satisfy the $F_4$ algebra,

\bea
[\,T^{}_{ij}\,,T^{}_{kl}\,]&=&-i\,(\delta^{}_{jk}\,T^{}_{il}+\delta^{}_{il}\,
T^{}_{jk}-\delta^{}_{ik}\,T^{}_{jl}-\delta^{}_{jl}\,T^{}_{ik})\ ,\\
~[\,T^{}_{ij}\,,T^{}_a\,]&=&\frac i2\,(\gamma^{}_{ij})^{}_{ab}\,T^{}_b\ ,\\
~[\,T^{}_a\,,T^{}_b\,]&=&\frac i2\,(\gamma^{}_{ij})^{}_{ab}\,T^{}_{ij}\ ,
\eea
so that the structure constants are given by

$$
f_{ij\,ab}~=~f_{ab\,ij}~=~\frac{1}{2}\,(\gamma^{}_{ij})_{ab}\ .$$

We can now use the generators $T^{}_a$ in Kostants equation and we 
can form the $\gamma$'s from the knowledge about $S^{}_{ij}$. The $\gamma$'s 
will be written in terms of the $\theta$'s 
and their derivatives. It is then straightforward but tedious to 
solve Kostant's eqation and we will obtain an infinite superfield 
written in terms of ordinary fields depending on $x$ expanded in 
$\theta$'s and creation operators of the oscillator algebra. For 
details see  \cite{Brink:2002zq}.

The next step to see is if also for this theory one can construct a 
three-point function along the lines of the corresponding work for the 
supergravity multiplet. Note that this theory is different from other 
approaches to find higher spin gauge theories in that it is built 
from triplets with very strong cancellations between the bosons and the 
fermions. If this is good enough to warrant a finite quantum theory 
rests to be seen.

\section{Afterthoughts}

In this talk I have tried to show how the idea of fermionic degress of 
freedom introduced via Grassmann odd coordinates has led us up to the 
Standard Model of particle physics and onto the Superstring Theory 
and perhaps further on. Some thirty years ago when we discussed 
fermionic coordinates we were very defensive and said that this is 
just a book-keeping device, but time has shown that it is more than 
that. The mathematics behind them is slightly different and we cannot 
grasp them like the three space dimensions around us. We were also 
ridiculed when we talked about $26$ or $10$ dimensions of space-time 
and also there we were defensive. Now the extra dimensions are taken 
for granted and in the same way as we have learnt to work with them we 
have learnt to work with the fermionic ones. I think this gives us a 
healthy perspective on the concepts of coordinates in general.

What would have been Gunnar Nordstr\"om's role had he not dribbled 
into research on radioactivity and forgotten uranium in his waist 
pocket? Had he been like a Sandels in Pardala by or a D\"obeln at 
Jutas leading the troups 
into the new field of quantum physics and new coordinates or more like the 
brave leutenant Zid\'{e}n storming along far in front of everybody 
\cite{Runeberg}. Unfortunately we cannot tell. 
Science lost one of its most inventive researchers when Gunnar 
Nordstr\"om died so young, and the developments of modern physics was slowed down. Even so the development is most impressive.

\end{document}